\documentstyle[12pt]{article}

\newcommand{\drawsquare}[2]{\hbox{%
\rule{#2pt}{#1pt}\hskip-#2pt
\rule{#1pt}{#2pt}\hskip-#1pt
\rule[#1pt]{#1pt}{#2pt}}\rule[#1pt]{#2pt}{#2pt}\hskip-#2pt%
\rule{#2pt}{#1pt}}

\newcommand{\fund}{\raisebox{-.5pt}{\drawsquare{6.5}{0.4}}}
\newcommand{\Ysymm}{\raisebox{-.5pt}{\drawsquare{6.5}{0.4}}\hskip-0.4pt%
        \raisebox{-.5pt}{\drawsquare{6.5}{0.4}}}
\newcommand{\Yasymm}{\raisebox{-3.5pt}{\drawsquare{6.5}{0.4}}\hskip-6.9pt%
        \raisebox{3pt}{\drawsquare{6.5}{0.4}}}
\newcommand{\antifund}{\overline{\fund}}

\def\yzero{\smash{\hbox{$y\kern-4pt\raise1pt\hbox{${}^\circ$}$}}}

\def\-{\hphantom{-}}
\def\ov{\overline}
\def\s2{\frac{1}{\sqrt2}}

\def\beq{\begin{equation}}
\def\eeq{\end{equation}}
\def\beqa{\begin{eqnarray}}
\def\eeqa{\end{eqnarray}}

\def\IF{\relax{\rm I\kern-.18em F}}
\def\II{\relax{\rm I\kern-.18em I}}
\def\IP{\relax{\rm I\kern-.18em P}}

\def\Dsl{\,\raise.15ex\hbox{/}\mkern-13.5mu D} 

\def\IC{\bf C}
\def\IZ{\bf Z}

\def\z2z2{$\IC^3/(\IZ_2\times\IZ_2)$}

\def\s{\sigma}

\def\z{\zeta}

\def\bo{{\raise-.3ex\hbox{\large$\Box$}}}

\def\face{{\raise.2ex\hbox{$\displaystyle \bigodot$}\mskip-2.2mu \llap
{$\ddot
        \smile$}}}

\def\leftrightarrowfill{$\mathsurround=0pt \mathord\leftarrow \mkern-6mu
        \cleaders\hbox{$\mkern-2mu \mathord- \mkern-2mu$}\hfill
        \mkern-6mu \mathord\rightarrow$}
\def\dvec#1{\vbox{\ialign{##\crcr
        \leftrightarrowfill\crcr\noalign{\kern-1pt\nointerlineskip}
        $\hfil\displaystyle{#1}\hfil$\crcr}}}           

\def\beq{\begin{equation}}
\def\eeq{\end{equation}}

\def\beqx{\begin{displaymath}}
\def\eeqx{\end{displaymath}}

\def\beqa{\begin{eqnarray}}
\def\eeqa{\end{eqnarray}}

\begin{document}
\input{psfig.sty}

\begin{flushright}
\baselineskip=12pt
UPR-1040-T \\
\end{flushright}

\begin{center}
\vglue 1.5cm

{\Large\bf Quasi-Supersymmetric $G^3$ Unification from
Intersecting D6-Branes
 on Type IIA Orientifolds}

\vglue 2.0cm

{\Large Tianjun Li$^a$\footnote{ E-mail: tli@sns.ias.edu. Phone
Number: (609) 734-8024. Fax Number:  (609) 951-4489.}
 and Tao Liu$^b$\footnote{E-mail: liutao@sas.upenn.edu. Phone Number:
(215) 573-6105.}} \vglue 1cm
{$^a$ School of Natural Science, Institute for Advanced Study,  \\
             Einstein Drive, Princeton, NJ 08540, USA\\
$^b$ Department of Physics and Astronomy, University of Pennsylvania, \\
Philadelphia, PA 19104, USA\\}

\end{center}

\vglue 1.5cm
\begin{abstract}
We construct three quasi-supersymmetric $G^3$ GUT models
with $S_3$ symmetry and gauge coupling
unification from intersecting D6-branes on Type IIA orientifolds.
The Standard Model fermions and Higgs doublets can be embedded
into the bifundamental representations in these models, and there
is no any other unnecessary massless representation. Especially in Model
I with gauge group $U(4)^3$, we just have three-family SM fermions
and three pairs of Higgs particles. The $G^3$ gauge symmetry in
these models can be broken down to the Standard Model gauge
symmetry by introducing light open string states. And 1
TeV scale supersymmetry breaking soft masses imply the reasonable
intermediate string scale.
\\[1ex]
\\[1ex]
\end{abstract}

\vspace{0.2cm}
\begin{flushleft}
\baselineskip=12pt
April 2003\\
\end{flushleft}
\newpage
\setcounter{page}{1} \pagestyle{plain} \baselineskip=14pt

\section{Introduction}

Since 1984, there has been a lot of work and effort devoted to the
string model building or string phenomenology, whose goal is to
obtain the Standard Model (SM) or Minimal Supersymmetric Standard
Model (MSSM) as an effective theory of the string-based models. And
these models are mainly built in the weakly coupled heterotic
string theory with $E_8\times E_8$ gauge group~\cite{GSW, JP},
because it naturally obtains the Grand Unified Theory (GUT)
through the elegant $E_8$ breaking chain: $E_8 \supset E_6 \supset
SO(10) \supset SU(5)$. Even now, this is an interesting subject
because of the model buildings in M-theory on $S^1/Z_2$ [3$-$7].

In recent years, the emergence of M-theory opened up many new
avenues for the consistent string model buildings. Especially, we
can construct the open string models that are non-perturbative
from the dual heterotic string description due to the advent of
D-branes~\cite{JPEW}. The technique of conformal field theory in
describing D-branes and orientifold planes on orbifolds has played
a key role in the construction of consistent 4-dimensional
supersymmetric $N=1$ chiral models on Type II orientifolds. There
are two kinds of theories which have chiral fermions from the
D-brane constructions: one from D-branes located at orbifold
singularities where the chiral fermions appear on the worldvolume
of D-branes [9$-$15] and the other one from intersecting D-branes
on Type II orientifold where the open string spectrum contains
chiral fermions localized at the D-brane intersections~\cite{bdl}.

For the second kind of scenarios, a lot of non-supersymmetric three-family
Standard-like models and GUT models were
explored in the begining~[17$-$30].  However,
 there are uncancelled Neveu-Schwarz-Neveu-Schwarz (NSNS)
tadpoles and may exist the gauge hierarchy problem. On the other
hand, since the first supersymmetric model with intersecting
D6-branes on $T^6/Z_2 \times Z_2$ was constructed in
Refs.~\cite{CSU1, CSU2}, the supersymmetric Standard-like models,
$SU(5)$ and Pati-Salam models have been discussed in detail
later~\cite{CPS, CP}, as well as the phenomenology~\cite{CLS1,
CLS2, CLW}. Moreover, the supersymmetric Pati-Salam models based
on $Z_4$ and $Z_4\times Z_2$ orientifolds with intersecting
D6-branes were also constructed~\cite{blumrecent, Honecker}. In
these models, the left-right symmetric gauge structure was
obtained by brane recombinations, so the final models do not have
the explicit toroidal orientifold construction, where the
conformal field theory can be applied for the calculation of the
full spectrum and couplings.

Looking back on these model buildings, we may find that people
took such philosophy: directly construct the familiar models, such
as Standard-like models, $SU(5)$ and Pati-Salam models, etc, from the
intersecting D-branes on type II orientifolds since these models
have been understood very well from the traditional phenomenological
analysis. Unfortunately, no GUT model with gauge coupling
unification has been built up due to the strong
constraint of RR-tadpole cancellation and supersymmetry (SUSY)
 preservation. In
this paper, we take a completely different philosophy:
constructing the ``natural'' 4-dimensional $N=1$ GUT models from
the intersecting D6-branes on Type IIA orientifolds where the
``natural'' means:

(1) Gauge coupling unification;

(2) The Standard Model gauge group is the subgroup of the gauge
symmetry at string scale, and three families of quarks and leptons
and a pair of the SM Higgs doublets are included in the massless
open string spectrum;

(3) The gauge symmetry at string scale can be broken down to the
Standard Model gauge symmetry via Higgs mechanism or Wilson
line;

(4) RR-tadpole cancellation. And the observable D6-branes preserve
the same 4-dimensional $N=1$ supersymmetry as the orbifold
 background.

Adding $S_3$ symmetry on the observable D6-branes and complex structure
moduli, we obtain three models with above four
properties from $T^6 /(Z_2 \times Z_2)$ orientifolds with
intersecting D6-branes. In these models, three stacks of physical
D6-branes, which form the observable sector,
 preserve the same 4-dimensional $N=1$ supersymmetry as the orbifold
background. To cancel the RR tadpole, we introduce one
stack of auxiliary D6-branes which wraps on the $\Omega R$ orientifold
 and has no intersection with three observable
D6-branes. However, the auxiliary D6-brane breaks above
 4-dimensional $N=1$ supersymmetry.
So, our model is quasi-supersymmetric~\footnote{In this paper,
the quasi-supersymmetry means that the observable D6-branes preserve
the same 4-dimensional $N=1$ supersymmetry as the orbifold
 background, which is broken by the auxiliary D6-brane.}, and there
may exist the uncancelled NSNS tadpoles. Concretely, Model I
describes $U(4)^3$ gauge theory with odd-family chiral fermion
spectrum, and Model II $U(4)^3$ gauge theory with even-family
chiral fermion spectrum, Model III $U(8)^3$ gauge theory with
even-family chiral fermion spectrum. In all these models, the
Standard Model fermions and Higgs particles are embedded into the
bifundamental representations, and the symmetric, anti-symmetric
or any other unnecessary massless representations are absent. In
particular, we just have three families of fermions and three
pairs of Higgs particles for Model I. We show that in model I the
$U(4)^3$ gauge symmetry can indeed be broken down to the Standard
Model gauge symmetry by introducing the light open string states,
and similar mechanism works for the Models II and III. Furthermore, we
discuss the supersymmetry breaking due to the auxiliary D6-brane,
and find that the 1 TeV scale soft masses imply the intermediate
string scale around $10^{11}\sim 10^{12}$ GeV, which is a
reasonable unification scale
for the Pati-Salam model~\cite{Goran} and can be
realized in large extra dimension scenario~\cite{ADD, AADD}.
However, the unification gauge coupling ($\alpha_{\rm GUT}$) is
seriously suppressed to $10^{-8}$, which implies the fine-tuning
in the RGE runnings of the gauge-couplings.

\section{Supersymmetric Model Buildings from
$T^6 /(Z_2 \times Z_2)$ Orientifolds with Intersecting D6-Branes}

In spite of non-supersymmetric essence of $G^3$ GUT models, 
the 4-dimensional $N=1$ supersymmetry are
required to be locally preserved in the observable sector in order
to solve the gauge hierarchy problem. So, we first review the rules to
construct the supersymmetric models from Type IIA orientifolds on
$T^6 /(Z_2 \times Z_2)$ with D6-branes at generic angles , and to
obtain the spectrum of massless open string states~\cite{CSU2}.
 Here, we follow the notation in Ref.~\cite{CPS}.

The starting point is Type IIA string theory compactified on a
$T^6 /(Z_2 \times Z_2)$ orientifold. We consider $T^{6}$ to be a
six-torus factorized as $T^{6} = T^{2} \times T^{2} \times T^{2}$
whose complex coordinates are $z_i$, $i=1,\; 2,\; 3$ for each of
the 2-torus, respectively. The $\theta$ and $\omega$
generators for the orbifold group
$Z_{2} \times Z_{2}$, which are associated
with their twist vectors $(1/2,-1/2,0)$ and $(0,1/2,-1/2)$
respectively, act on the complex coordinates of $T^6$ as
\beqa
& \theta: & (z_1,z_2,z_3) \to (-z_1,-z_2,z_3)~,~ \nonumber \\
& \omega: & (z_1,z_2,z_3) \to (z_1,-z_2,-z_3)~.~\,
\label{orbifold}
\eeqa
The orientifold projection is implemented
by gauging the symmetry $\Omega R$, where $\Omega$ is world-sheet
parity, and $R$ acts as 
\beqa
 R: (z_1,z_2,z_3) \to ({\ov z}_1,{\ov z}_2,{\ov
z}_3)~.~\, \label{orientifold} 
\eeqa 
So, there are four kinds of
orientifold 6-planes (O6-planes) for the actions of $\Omega R$,
$\Omega R\theta$, $\Omega R \omega$, and $\Omega R\theta\omega$,
respectively. To cancel the RR charges of O6-planes, we introduce
some stacks of $N_a$ D6-branes, which wrap on the factorized
three-cycles. Meanwhile, we have two kinds of complex structures
consistent with orientifold projection for a torus -- rectangular
and tilted~\cite{bkl, CSU2, CPS}. If we denote the homology classes of
the three cycles wrapped by the D6-brane stacks as
$n_a^i[a_i]+m_a^i[b_i]$ and $n_a^i[a'_i]+m_a^i[b_i]$ with
$[a_i']=[a_i]+\frac{1}{2}[b_i]$ for the rectangular and tilted
tori respectively, following the notation of Ref.~\cite{CPS}, we
can label a generic two cycle by $(n_a^i,l_a^i)$ in either case,
where in terms of the wrapping numbers $l_{a}^{i}\equiv m_{a}^{i}$
for a rectangular torus and $l_{a}^{i}\equiv
2\tilde{m}_{a}^{i}=2m_{a}^{i}+n_{a}^{i}$ for a tilted torus. Note
that for a tilted torus, $l_a^i-n_a^i$ must be even. For a stack
of $N_a$ D6-branes along the cycle $(n_a^i,l_a^i)$, we also need
to include their $\Omega R$ images $N_{a'}$ with wrapping numbers
$(n_a^i,-l_a^i)$. For D6-branes on the top of O6-planes, we count
the D6-branes and their images independently. So, the homology
three-cycles for stack $a$ of $N_a$ D6-branes and its orientifold
image $a'$ take the form
\beq
[\Pi_a]=\prod_{i=1}^{3}\left(n_{a}^{i}[a_i]+2^{-\beta_i}l_{a}^{i}[b_i]\right),\;\;\;
\left[\Pi_{a'}\right]=\prod_{i=1}^{3}
\left(n_{a}^{i}[a_i]-2^{-\beta_i}l_{a}^{i}[b_i]\right)~,~\,
\eeq
where $\beta_i=0$ if the $i-th$ torus is rectangular and
$\beta_i=1$ if it is tilted. And the homology 3-cycles wrapped by
the four O6-planes are
\beq
\Omega R: [\Pi_{\Omega R}]= 2^3
[a_1]\times[a_2]\times[a_3]~,~\, \eeq \beq \Omega R\omega:
[\Pi_{\Omega
R\omega}]=-2^{3-\beta_2-\beta_3}[a_1]\times[b_2]\times[b_3]~,~\,
\eeq \beq \Omega R\theta\omega: [\Pi_{\Omega
R\theta\omega}]=-2^{3-\beta_1-\beta_3}[b_1]\times[a_2]\times[b_3]~,~\,
\eeq
\beq
 \Omega R\theta:  [\Pi_{\Omega
R}]=-2^{3-\beta_1-\beta_2}[b_1]\times[b_2]\times[a_3]~.~\,
\label{orienticycles} \eeq Then, the intersection numbers are \beq
I_{ab}=[\Pi_a][\Pi_b]=2^{-k}\prod_{i=1}^3(n_a^il_b^i-n_b^il_a^i)~,~\,
\eeq
\beq
I_{ab'}=[\Pi_a]\left[\Pi_{b'}\right]=-2^{-k}\prod_{i=1}^3(n_{a}^il_b^i+n_b^il_a^i)~,~\,
\eeq
\beq
I_{aa'}=[\Pi_a]\left[\Pi_{a'}\right]=-2^{3-k}\prod_{i=1}^3(n_a^il_a^i)~,~\,
\eeq
\beq {I_{aO6}=[\Pi_a][\Pi_{O6}]=2^{3-k}(-l_a^1l_a^2l_a^3
+l_a^1n_a^2n_a^3+n_a^1l_a^2n_a^3+n_a^1n_a^2l_a^3)}~,~\,
\label{intersections}
\eeq
where $[\Pi_{O6}]=[\Pi_{\Omega
R}]+[\Pi_{\Omega R\omega}]+[\Pi_{\Omega
R\theta\omega}]+[\Pi_{\Omega R\theta}]$ is the sum of O6-plane
homology three-cycles wrapped by the four O6-planes,
 and $k=\beta_1+\beta_2+\beta_3$ is
the total number of tilted tori.

\begin{table}[t]
\caption{General spectrum on intersecting D6-branes at generic
angles which is valid for both rectangular and tilted tori. The
representations in the table make sense to $U(N_a/2)$ due to
$Z_2\times Z_2$ orbifold projection~\cite{CSU2}. In supersymmetric
situations, scalars combine with the fermions to form the chiral
supermultiplets.}
\renewcommand{\arraystretch}{1.25}
\begin{center}
\begin{tabular}{|c|c|}
\hline {\bf Sector} & \phantom{more space inside this box}{\bf
Representation}
\phantom{more space inside this box} \\
\hline\hline
$aa$   & $U(N_a/2)$ vector multiplet  \\
       & 3 Adj. chiral multiplets  \\
\hline
$ab+ba$   & $I_{ab}$ $(\fund_a,\antifund_b)$ fermions   \\
\hline
$ab'+b'a$ & $I_{ab'}$ $(\fund_a,\fund_b)$ fermions \\
\hline $aa'+a'a$ &$-\frac 12 (I_{aa'} - \frac 12 I_{a,O6})\;\;
\Ysymm\;\;$ fermions \\
          & $-\frac 12 (I_{aa'} + \frac 12 I_{a,O6}) \;\;
\Yasymm\;\;$ fermions \\
\hline
\end{tabular}
\end{center}
\label{spectrum}
\end{table}

The general spectrum on intersecting D6-branes at generic angles,
which is valid for both rectangular and tilted tori, is given in
Table \ref{spectrum}. And the 4-dimensional chiral supersymmetric
(N=1) models from Type IIA Orientifolds with intersecting
D6-branes are mainly constrained in two
aspects:\\

I. Tadpole Cancellation Conditions\\

As sources of RR fields, D6-branes and orientifold 6-planes are
required to satisfy the Gauss law in a compact space, {\it i.e.},
the total RR charges of D6-branes and O6-planes must vanish since
the RR field flux lines can't escape.
 The RR tadpole cancellation conditions are
\begin{eqnarray}
\sum_a N_a [\Pi_a]+\sum_a N_a
\left[\Pi_{a'}\right]-4[\Pi_{O6}]=0~,~\,
\end{eqnarray}
where the last contributions come from the O6-planes which have $-4$
RR charges in the D6-brane charge unit by exchanging RR field
while scattering.

Tadpole cancellation directly leads to the $SU(N)^3$ cubic
non-abelian anomaly cancellation~\cite{Uranga, imr, CSU2}. And the
cancellation of U(1) mixed gauge and gravitational anomaly or
$[SU(N)]^2 U(1)$ gauge anomaly can be achieved by Green-Schwarz
mechanism mediated by untwisted RR fields~\cite{Uranga, imr,
CSU2}.\\

\renewcommand{\arraystretch}{1.4}
\begin{table}[t]
\caption{Wrapping numbers of the four O6-planes.} \vspace{0.4cm}
\begin{center}
\begin{tabular}{|c|c|c|}
\hline
  Orientifold Action & O6-Plane & $(n^1,l^1)\times (n^2,l^2)\times
(n^3,l^3)$\\
\hline
    $\Omega R$& 1 & $(2^{\beta_1},0)\times (2^{\beta_2},0)\times
(2^{\beta_3},0)$ \\
\hline
    $\Omega R\omega$& 2& $(2^{\beta_1},0)\times (0,-2^{\beta_2})\times
(0,2^{\beta_3})$ \\
\hline
    $\Omega R\theta\omega$& 3 & $(0,-2^{\beta_1})\times
(2^{\beta_2},0)\times
(0,2^{\beta_3})$ \\
\hline
    $\Omega R\theta$& 4 & $(0,-2^{\beta_1})\times (0,2^{\beta_2})\times
    (2^{\beta_3},0)$ \\
\hline
\end{tabular}
\end{center}
\label{orientifold}
\end{table}

II. Conditions for 4-dimensional $N = 1$ Supersymmetric D6-brane \\

The 4-dimensional $N=1$ supersymmetric models require that 1/4
supercharges from 10-dimensional Type I T-dual be preserved, {\it
i.e}, they should survive two supersymmetry breaking
mechanisms: orientation projection of the intersecting D6-branes,
and orbifold projection on the background manifold.
Concrete analysis shows that the $N=1$ supersymmetry can be
preserved only if the rotation angle of any D6-brane with respect
to the $\Omega R$-plane is an element of $SU(3)$, or in other
words, $\theta_1+\theta_2+\theta_3=0 $, where $\theta_i$ is the
angle between the $D6$-brane and the $\Omega R$-plane in the
$i-th$ torus. In Ref.~\cite{CPS}, this condition is rewritten as
\begin{eqnarray}
 -x_A l_a^1l_a^2l_a^3+x_B
l_a^1n_a^2n_a^3+x_C n_a^1l_a^2n_a^3+x_D n_a^1n_a^2l_a^3=0~,~\,
\label{susycondition1}
\end{eqnarray}
\begin{eqnarray}
-n_a^1n_a^2n_a^3/x_A+n_a^1l_a^2l_a^3/x_B+l_a^1n_a^2l_a^3/x_C+l_a^1l_a^2n_a^3/x_D<0~,~\,
\label{susycondition2}
\end{eqnarray}
where $x_A=\lambda,\; x_B=\lambda
2^{\beta_2+\beta3}/\chi_2\chi_3,\; x_C=\lambda
2^{\beta_1+\beta3}/\chi_1\chi_3,\; x_D=\lambda
2^{\beta_1+\beta2}/\chi_1\chi_2$, and $\chi_i=R_2^i/R_1^i$ are the
complex structure moduli where $R_1^i$ and $R_2^i$ are radii for
the $i-th$ torus due to $T^2\equiv S^1\times S^1$. $\lambda$ is a
positive parameter without physical significance.

\section{Quasi-Supersymmetric $G^3$ Unification}
Generally speaking, the RR tadpole cancellation conditions and the
 4-dimensional supersymmetry
 preservation conditions are too stringent to find the
realistic GUT models, and the existing GUT models always tend to
produce extra gauge interactions and extra fermions beyond the SM
or MSSM. However, by relaxing the supersymmetry preserving condition for
the auxiliary D6-brane which is introduced to cancel the RR
tadpole, we can construct the natural GUT models with the
four properties emphasized in Introduction.

Let us look at the tadpole cancellation conditions first. If we
consider $N^{(i)}$ auxiliary D6-branes wrapped along the $i-th$ orientifold
plane whose wrapping numbers are given in Table \ref{orientifold},
the tadpole cancellation conditions are modified to
\beqa
 -2^k N^{(1)} - \sum_{\sigma} N_{\sigma}
 n_{\sigma}^1 n_{\sigma}^2 n_{\sigma}^3 = -16~,~\,
\label{tad4} \eeqa \beqa -2^k N^{(2)} + \sum_{\sigma} N_{\sigma}
n_{\sigma}^1 l_{\sigma}^2 l_{\sigma}^3 =-16~,~\, \label{tad1}
\eeqa \beqa -2^k N^{(3)}+\sum_{\sigma} N_{\sigma} l_{\sigma}^1
n_{\sigma}^2 l_{\sigma}^3 =-16~,~\, \label{tad2} \eeqa \beqa -2^k
N^{(4)}+\sum_{\sigma} N_{\sigma} l_{\sigma}^1 l_{\sigma}^2
n_{\sigma}^3=-16 ~.~\, \label{tad3}
\eeqa
Suppose there are three stacks of observable D6-branes, $a$, $b$, and
$c$. Adding $S_3$ symmetry
onto D6-branes configuration and $T_2 \times T_2 \times T_2$
geometry, {\it i.e.}, $N^{(2)}=N^{(3)}=N^{(4)}$, $N_a=N_b=N_c=2N$ and
$\chi_1=\chi_2=\chi_3$, we notice that among Eqs. (\ref{tad1}),
(\ref{tad2}) and (\ref{tad3}), only one is independent.
Similarly for the
$N=1$ supersymmetry preserving conditions. If
 one stack of the observable D6-brane preserves $N=1$ supersymmetry, all three
stacks of D6-branes will preserve the $N=1$ supersymmetry
automatically. The simplest case is that
$N^{(2)}=N^{(3)}=N^{(4)}=0$, and one stack of auxiliary D6-brane
wrapped along the
 $\Omega R$ orientifold plane are needed for RR tadpole
cancellation in these models. Then the gauge group of our models is
$G^3$ where $G=U(N)$.

For simplicity, we consider three stacks of observable D6-branes ($a$, $b$
and $c$) with one zero wrapping number. Without loss of
generality, we have two posssibilities
\begin{eqnarray}
n_{a}^{1}=n_{b}^{2}=n_{c}^{3}=0 ~(i)~;~
l_{a}^{1}=l_{b}^{2}=l_{c}^{3}=0 ~(ii)~.~\,
\end{eqnarray}
For the first case ($i$), the models without symmetric and antisymmetric
representations can not be constructed. So, we focus on the second
case ($ii$).

\renewcommand{\arraystretch}{1.4}
\begin{table}[t]
\caption{Model I. D6-brane configuration in (2p+1)-generation
quasi-supersymmetric $U(4)^3$ model. This model is built on three
tilted 2-tori with $Z_2\times Z_2$ orbifold symmetry and p is a
non-negative integer.} \vspace{0.4cm}
\begin{center}
\begin{tabular}{|c|c|c|c|}
\hline $N_i$ & $(n_{i}^{1}, l_{i}^{1})$ & $(n_{i}^{2}, l_{i}^{2})$
& $(n_{i}^{3},
l_{i}^{3})$ \\
\hline
$N_a=8$ & $(2,0)$ & $(2p+1,1)$ & $(2p+1,-1)$ \\
\hline
$N_b=8$ & $(2p+1,-1)$ & $(2,0)$ & $(2p+1,1)$ \\
\hline
$N_c=8$ & $(2p+1,1)$ & $(2p+1,-1)$ & $(2,0)$ \\
\hline $N_g$ & \multicolumn{3}{c|}{$N_g n_g^1 n_g^2
n_g^3=-48(2p+1)^2+16$} \\
\hline
\end{tabular}
\end{center}
\label{sol1}
\end{table}

\renewcommand{\arraystretch}{1.4}
\begin{table}[t]
\caption{Model II. D6-brane configuration in (8p)-generation
quasi-supersymmetric $U(4)^3$ model. This model is built on three
rectangular 2-tori with $Z_2\times Z_2$ orbifold symmetry and p is
a positive integer.} \vspace{0.4cm}
\begin{center}
\begin{tabular}{|c|c|c|c|}
\hline $N_i$ & $(n_{i}^{1}, l_{i}^{1})$ & $(n_{i}^{2}, l_{i}^{2})$
& $(n_{i}^{3},
l_{i}^{3})$ \\
\hline
$N_a=8$ & $(2,0)$ & $(p,1)$ & $(p,-1)$ \\
\hline
$N_b=8$ & $(p,-1)$ & $(2,0)$ & $(p,1)$ \\
\hline
$N_c=8$ & $(p,1)$ & $(p,-1)$ & $(2,0)$ \\
\hline
$N_g$ & \multicolumn{3}{c|}{$N_g n_g^1 n_g^2 n_g^3=-48p^2+16$} \\
\hline
\end{tabular}
\end{center}
\label{sol2}
\end{table}

\renewcommand{\arraystretch}{1.4}
\begin{table}[t]
\caption{Model III. D6-brane configuration in (2p)-generation
quasi-supersymmetric $U(8)^3$ model. This model is built on three
rectangular 2-tori with $Z_2\times Z_2$ orbifold symmetry and p is
a positive integer.} \vspace{0.4cm}
\begin{center}
\begin{tabular}{|c|c|c|c|}
\hline $N_i$ & $(n_{i}^{1}, l_{i}^{1})$ & $(n_{i}^{2}, l_{i}^{2})$
& $(n_{i}^{3},
l_{i}^{3})$ \\
\hline
$N_a=16$ & $(1,0)$ & $(p,1)$ & $(p,-1)$ \\
\hline
$N_b=16$ & $(p,-1)$ & $(1,0)$ & $(p,1)$ \\
\hline
$N_c=16$ & $(p,1)$ & $(p,-1)$ & $(1,0)$ \\
\hline
$N_g$ & \multicolumn{3}{c|}{$N_g n_g^1 n_g^2 n_g^3=-48p^2+16$} \\
\hline
\end{tabular}
\end{center}
\label{sol3}
\end{table}

\renewcommand{\arraystretch}{1.4}
\begin{table}[t]
\caption{Chiral open string spectrum for the $U(N)^3$ GUT models.
$N=4$ for Model I and Model II, and $N=8$ for Model III. $N_f =
2p+1, ~8p, ~2p$ for Model I, Model II, and Model III,
respectively.} \vspace{0.4cm}
\begin{center}
\begin{tabular}{|c||c||c|c|c|}
\hline Sector & $U(N) \times U(N) \times U(N)$ &
$Q_a$ & $Q_b$ & $Q_c$ \\
\hline
$ab + ba$ & $N_f \times (N,{\ov N},1)$ & $1$ & $-1$ & $0$ \\
\hline
$bc + cb$ & $N_f \times (1,N,{\ov N})$ & $0$ & $1$ & $-1$ \\
\hline
$ca + ac$ & $N_f \times ({\ov N},1,N)$ & $-1$ & $0$ & $1$ \\
\hline
\end{tabular}
\end{center}
\label{spectrum4}
\end{table}

In addition, we only consider the models with bifundamental
representations which the Standard Model fermions and Higgs 
particles can be
embedded into. To avoid the symmetric and anti-symmetric
representations, we require that
\begin{eqnarray}
l_{a}^{2}n_{a}^{3}=-n_{a}^{2}l_{a}^{3}~;~ &
l_{b}^{3}n_{b}^{1}=-n_{b}^{3}l_{b}^{1}~;~ &
l_{c}^{1}n_{c}^{2}=-n_{c}^{1}l_{c}^{2}~,~\, \label{vanish}
\end{eqnarray}
which are equivalent to the supersymmetry preserving conditions.
Because of the $S_3$ symmetry among the three stacks of D6-branes
or three 2-tori, Eq. (\ref{vanish}) implies
\begin{eqnarray}
l_{b}^{3}n_{a}^{3}=-n_{b}^{3}l_{a}^{3}~;~ &
l_{c}^{1}n_{b}^{1}=-n_{c}^{1}l_{b}^{1}~;~ &
l_{a}^{2}n_{c}^{2}=-n_{a}^{2}l_{c}^{2}~,~\,
\end{eqnarray}
and vice versa. This means that at massless level, the
representations $(N_a/2, N_b/2, 1)$,
 $(1,N_b/2,N_c/2)$, $(N_a/2, 1, N_c/2)$ (or their complex conjugations)
will appear or disappear together with the symmetric and
anti-symmetric representations in the models with $G^3$
unification. As for the determination of $N$ in $U(N)^3$ gauge
group, we only have two choices: 4 or 8, which can be figured out
from the tadpole cancellation conditions in our setup:
\begin{eqnarray}
N_a n_{a}^{1}l_{a}^{2}l_{a}^{3}=-16~,~ N_b
l_{b}^{1}n_{b}^{2}l_{b}^{3}=-16~,~ N_c
l_{c}^{1}l_{c}^{2}n_{c}^{3}=-16~,~ \label{tadpole1}
\end{eqnarray}
\begin{eqnarray}
-(N_a n_{a}^{1}n_{a}^{2}n_{a}^{3}+N_b n_{b}^{1}n_{b}^{2}n_{b}^{3}
+N_cn_{c}^{1}n_{c}^{2}n_{c}^{3})-N_{g}n_{g}^{1}n_{g}^{2}n_{g}^{3}=-16.
\label{tadpole2}
\end{eqnarray}
where $N_a=N_b=N_c=2N$. Obviously, $N$ can't be larger than 8
since the four O6-planes in our setup can only provide $-16$ RR
charges in the D6-brane charge unit, while $N=2$ is ruled out from
 the phenomenological concern. We emphasize that for $U(4)^3$
model, the three tori can be tilted, but, for $U(8)^3$ model, the
three tori can not be tilted since $n_a^1-l_a^1$ is odd.

There are three typical solutions corresponding to three $G^3$
models. The D6-brane configurations for Model I, Model II, and
Model III are given in Tables \ref{sol1}, \ref{sol2}, and
\ref{sol3}, respectively. We also present the chiral open string
spectrum for those models in Table \ref{spectrum4}. In short, we
have $2p+1$, $8p$ and $2p$ generations of bifundamental
representations under $U(N)^3$ gauge symmetry which include the
Standard Model fermions and Higgs particles. In particular, in
Model I, we can only have three families of fermions and three
pairs of Higgs particles.

One may notice that in Tables \ref{sol1}, \ref{sol2}, and
\ref{sol3}, the number of the auxiliary branes ($N_g$) is negative if
we have at least three family fermions. This means that the
auxiliary branes are anti-D6-branes. And then, the 4-dimensional
$N=1$ supersymmetry, which is preserved by the observable D6-branes
and orbifold background, is broken by the auxiliary D6-branes. Therefore,
the models are quasi-supersymmetric, and the NSNS tadpoles do not vanish.

\section{Comments on Phenomenology of $G^3$ Models}

\subsection{Gauge Coupling Unification}

The gauge couplings have been discussed in Refs.~\cite{CLS1, CLW}.
Since the gauge couplings are associated with different stacks of
D6-branes, usually they do not have a conventional gauge coupling
unification, although the value of each gauge coupling at the
string scale is predicted in terms of the moduli $\chi_i$ and the
ratio of the Planck scale to string scale. Let us calculate the
4-dimensional gauge coupling in detail, and show that in our
models, we do have the gauge coupling unification.

Dp-branes provide us a world where the gauge sectors are
localized on $(p+1)$-dimensional space-time
while gravity propagates in 10-dimensional space-time.
 Before compactification, the
gravitational and gauge interaction on Dp-brane can be generally
described by an effective action~\cite{CVJN}
\beqa S_{10}
\supset
\int \, d^{10}x \, \frac{M_s^{\, 8}}{(2\pi)^7 g_s^2} \, R_{10d} \,
+ \int \, d^{p+1}x \, \frac{M_s^{p-3}}{(2\pi)^{p-2} g_s} \,
F_{p+1}^{\, 2}~,~
\eeqa
where $M_s=1/\sqrt{\alpha'}$ is the string
scale, and $g_s$ is the string coupling.
 Upon the compactification, the 4-dimensional Planck scale $M_{Pl}$
 and the gauge coupling $g_{YM}^{\sigma}$ on the D6-brane stack $\sigma$ are
\beqa
 M_{Pl}^2 = \frac{ M_s^8 V_6
}{ (2 \pi)^7 g_s^2} ~,~ (g_{YM}^{\sigma})^2 =  \frac{(2 \pi)^4
g_s}{ M_s^3 V_3^{\sigma}}~.~\, \eeqa where \beqa
 V_6 = \frac{ (2 \pi)^6}{4}
\prod_{i=1}^3 R_1^i R_2^i~,~\,\label{mpms} 
\eeqa 
is the physical
volume of $T^6$ and 
\beqa
 V_3^{\sigma} = {1\over 4} (2 \pi)^3 \prod_{i=1}^3
\sqrt{\left(n_{\sigma}^{i} R_1^i\right)^2 + \left(2^{-\beta_i}
l_{\sigma}^i R_2^i\right)^2}~,~\, \label{v3} 
\eeqa 
is the physical
volume of three-cycle wrapped by the D6-brane stack $\sigma$. 
So, we obtain
\beqa
(g_{YM}^{\sigma})^2 = {{{\sqrt {8\pi}} M_s}\over\displaystyle
{M_{Pl}}} {1\over\displaystyle {\prod_{i=1}^3 {\sqrt
{\left(n_{\sigma}^{i} \right)^2\chi_i^{-1} + \left(2^{-\beta_i}
l_{\sigma}^i \right)^2 \chi_i}}}}~.~\, 
\label{coupling}
\eeqa 
Because in our models,  $\chi_1= \chi_2= \chi_3= \chi$,
we do have the gauge coupling unification. In general, we can expect
that $n_{\sigma}^{i}$, $l_{\sigma}^i$ and $\chi_i$ are the order
one integer or real number. Then the 4-dimensional gauge coupling
$(g_{YM}^{\sigma})^2$ is about ${{M_s}/{M_{Pl}}}$.
Therefore, for the intersecting D6-brane models with low string scale
on the space-time $M^4\times T^6$
or $M^4\times T^6/(Z_2\times Z_2)$,
where the D6-branes wrap on the factorized three cycles
of three 2-tori, the gauge couplings are generically
very small and may lead to the fine-tuning
 in the RGE runnings of gauge couplings.
However, in the general Calabi-Yau threefolds, 
one can make the physical volume of the 6-dimensional
 compact manifold large without affecting the physical 
volume of the compact three cycles wrapped by the 
D6-branes~\cite{BBKLU1, BBKLU2},
so, the low string scale in D6-brane models
 does not imply the very small
gauge couplings in general.

\subsection{Gauge Symmetry Breaking}

In our models, the $U(N)^3$ gauge symmetry can be broken down to
the Standard Model gauge symmetry by introducing the light open
string states. As an example, we only consider the Model I, and
similarly, one can discuss the gauge symmetry breaking in Model II
and Model III.

In Model I, we have 3 families by choosing $p=1$. The gauge group
is $U(4)\times U(4)\times U(4)$, which has subgroup $SU(4)\times
SU(2) \times SU(2)$, {\it i.e.}, the Pati-Salam model. The
left-handed fermions come from the $(4, \bar 4, 1)$
representations, the right-handed fermions come from the $(\bar 4,
1, 4)$ representations, and the pair of Higgs doublets come from
the $(1, 4, \bar 4)$ representations. Then, we will have three
pairs of Higgs doublets. However, in order to have the D-flat and
F-flat directions, we find that there are no Higgs particles at
massless state level which can break the $U(4)\times U(4)\times
U(4)$ gauge symmetry down to the $SU(4)\times SU(2) \times SU(2)$
or Standard Model gauge symmetry. Thus, the GUT breaking Higgs
fields must arise from the light open string spectrum.

Indeed, we do have such kind of Higgs fields. The ``$a$'' stack of
D6-branes $a$ is parallel to the orientifold ($\Omega R$) image
$b'$ of the ``$b$'' stack of D6-branes along the third torus, {\it
i.e.},
 the ``$b$'' stack of D6-branes $b$ is parallel to the orientifold
($\Omega R$) image $a'$ of the ``$a$'' stack of D6-branes along
the third torus. Then, there are open strings which stretch
between the branes $a$ and $b'$ (or say $a'$ and $b$). If the
minimal distance squared $Z^2_{(ab')}$
 (in $\alpha'$ units) between these two branes on the third torus is
small, {\it i.e.},  the minimal length squared of the stretched
string is small, we have the light scalars with masses
$Z^2_{(ab')}/(4\pi^2 \alpha')$ from the NS sector, and the light
fermions with the same masses from the R sector~\cite{Uranga,
imr}. These scalars and fermions form the 4-dimensional $N=2$
hypermultiplets. Similarly, the ``$b$'' stack of D6-branes $b$ is
parallel to the orientifold ($\Omega R$) image $c'$ of the ``$c$''
stack of D6-branes along the first torus, and the ``$c$'' stack of
D6-branes $c$ is parallel to the orientifold ($\Omega R$) image
$a'$ of the ``$a$'' stack of D6-branes along the second torus.
Thus, we can also have the light hypermultiplets from the open
strings which stretch between the branes $b$ and $c'$, and
between the branes $c$ and $a'$.

The light open string spectrum is given in Table \ref{spectrum5}.
These light Higgs fields can break the $U(4)^3$ down to the
Standard Model gauge symmetry. Roughly speaking, the Higgs fields
in the $(1,4,4)$ and $(1,{\ov 4}, {\ov 4})$ representations can
break the $U(4)\times U(4)\times U(4)$ gauge symmetry down to the
$U(4)\times SU(2) \times SU(2)$ gauge symmetry, and the Higgs
fields in the $(4, 1, 4)$ and $({\ov 4}, 1, {\ov 4})$
representations can break the $U(4)\times SU(2) \times SU(2)$
gauge symmetry down to the Standard Model gauge symmetry. The
detail symmetry breaking pattern and phenomenology are under
investigation. By the way, we do not need the particles in the
$(4,4,1)$ and $({\ov 4},{\ov 4},1)$ representations to be light
because we do not need them to break the gauge symmetry.

\renewcommand{\arraystretch}{1.4}
\begin{table}[t]
\caption{Light open string spectrum in the Model I which can break
the $U(4)^3$ gauge symmetry down to the Standard Model gauge
symmetry.} \vspace{0.4cm}
\begin{center}
\begin{tabular}{|c||c||c|c|c|c|}
\hline Sector & $U(N) \times U(N) \times U(N)$ &
$Q_a$ & $Q_b$ & $Q_c$ & Mass Square\\
\hline $ab' + ba'$ & $4 \times (4,4,1)$ & $1$ & $1$ & $0$ &
$Z^2_{(ab')}/(4\pi^2
\alpha')$ \\
$ab' + ba'$ & $4 \times ({\ov 4},{\ov 4},1)$ & $-1$ & $-1$ & $0$
 & $Z^2_{(ab')}/(4\pi^2 \alpha')$ \\
\hline $bc' + cb'$ & $4 \times (1,4,4)$ & $0$ & $1$ & $1$ &
$Z^2_{(bc')}/(4\pi^2
\alpha')$ \\
$bc' + cb'$ & $4 \times (1,{\ov 4}, {\ov 4})$ & $0$ & $-1$ & $-1$
&
 $Z^2_{(bc')}/(4\pi^2 \alpha')$ \\
\hline $ca' + ac'$ & $4 \times (4, 1, 4)$ & $1$ & $0$ & $1$ &
$Z^2_{(ca')}/(4\pi^2 \alpha')$ \\
$ca' + ac'$ & $4 \times ({\ov 4}, 1, {\ov 4})$ & $-1$ & $0$ & $-1$
& $Z^2_{(ca')}/(4\pi^2 \alpha')$ \\
\hline
\end{tabular}
\end{center}
\label{spectrum5}
\end{table}

\subsection{Supersymmetry Breaking and Possible Problems}

In our models, the observable D6-branes preserve the same
4-dimensional $N=1$ supersymmetry as the orbifold background does.
But, this supersymmetry is broken by the auxiliary D6-brane, which has
no interesections with the observable D6-branes.
So, the supersymmetry breaking effects can be mediated by
the heavy bifundamental messenger fields with string scale masses
which are the open strings
strectching between the observable D6-brane and auxiliary D6-brane,
and by the gravity supermultiplets in the bulk. Of course,
the dominant contributions to the scalar masses and gaugino masses
are from the gauge mediated supersymmetry breaking.

Similar to the discussions in~\cite{CIM1, CIM2}, the quadratic
divergences for scalars (for example Higgs fields) are absent up
to one-loop. The supersymmetry breaking soft masses
for scalars generated from two-loop diagrams are the same order  
as the gaugino masses generated from one-loop diagrams.
The soft masses-squared
for scalars $\phi_a$ typically are
\begin{eqnarray}
        {\tilde m}^2_a \propto [\frac{\alpha_i}{4\pi}]^2~M_s^2~.~\,
\end{eqnarray}
In our models,  $\chi_1= \chi_2= \chi_3=\chi$. Using Eq. (\ref{coupling}),  we obtain 
\beqa
M_s^2 \sim {{4\pi^2}\over {\sqrt {8 \pi}}} {\tilde m}_a M_{Pl}~
 {\prod_{i=1}^3 {\sqrt
{\left(n_{\sigma}^{i} \right)^2\chi^{-1} + \left(2^{-\beta_i}
l_{\sigma}^i \right)^2 \chi}}}~.~\, 
\eeqa 
Considering the Model I with three families and $\chi=1$, we obtain 
that the string scale $M_s$ is about $5.6\times 10^{11}$ GeV if
${\tilde m}_a \sim 1$ TeV. This is a
reasonable unification scale for the Pati-Salam model~\cite{Goran} and
can be generated by introducing relatively large extra
dimension~\cite{ADD, AADD} or small string coupling $g_s$.
However, the gauge coupling ($\alpha_{\rm GUT}$) at string scale
is seriously suppressed to $10^{-8}$, which implies the
fine-tuning in the RGE runnings of gauge coulings. For the RGE
runnings of gauge couplings, we should include the additional
contributions from the extra adjoint fields and their KK modes,
and the KK modes of gauge fields. Whether we can have such small
gauge coupling at string scale is a question deserving further
detail study. By the way, if $\chi$, which is a positive real number,
 is larger or smaller than 1, we can increase the string scale.
However, the unification gauge coupling at string scale is the same.

\section{Discussions and Conclusions}
Adding $S_3$ symmetry onto the observable D6-brane configuration
and complex structure moduli, we obtain three natural
quasi-supersymmetric GUT models with four interesting properties.
In Model I and Model II, the gauge group is $U(4)^3$, while in
Model III the gauge group is $U(8)^3$. The three tori of $T^6$ are
all tilted for Model I, and they are all rectangular for Model II
and Model III. The D6-brane configurations and chiral open string
spectrum at massless level are given in Tables
\ref{sol1}$-$\ref{spectrum4}. In all our three models, the
Standard Model fermions and Higgs particles can be embedded into
the bifundamental representations, and there is no any other unnecessary
massless representations. In particular, we only have three
families of fermions and three pairs of Higgs particles for Model
I. Moreover,
 we show that there exists the gauge coupling unification in
our models. We consider the gauge symmetry breaking, too.
Explicitly, we show that in Model I, the $U(4)\times U(4) \times
U(4)$ gauge symmetry can indeed be broken down to the Standard
Model gauge symmetry by introducing the light open string states,
and similar mechanism works for the Models II and III. Furthermore,
we find that the 1 TeV scale soft masses imply the intermediate
string scale ($10^{11}\sim 10^{12}$ GeV), which is a reasonable 
unification scale for
the Pati-Salam model. However, the unification gauge
coupling at string scale is very small and may lead to the
fine-tuning in the RGE runnings of gauge coulings. 

\section*{Acknowledgments}
We would like to thank Fernando Marchesano and Ioannis
Papadimitriou for helpful discussions, and thank Huiyu Albert Li
for inspirational conversations. The research of T. Li was
supported by the National
 Science Foundation under Grant No.~PHY-0070928.
 And the research of T. Liu was supported in part
 by the U.S.~Department of Energy under Grant
 No.~DOE-EY-76-02-3071.

\end{document}